\documentclass[pra,
 reprint,
superscriptaddress,
 amsmath,amssymb,
 aps,longbibliography
]{revtex4-1}

\usepackage{graphicx}% Include figure files
\usepackage{dcolumn}% Align table columns on decimal point
\usepackage{bm}% bold math
\usepackage{amsmath,amssymb,amstext,amsthm}
\usepackage{braket}
\usepackage{bbold}
\usepackage{bbm}
\usepackage{xcolor}
\usepackage{ gensymb }
\usepackage{soul}
\setstcolor{red}
\usepackage{ mathrsfs }
\usepackage{cancel}

\usepackage{hyperref}
\hypersetup{colorlinks,linkcolor=blue,citecolor=blue,urlcolor=blue}

\newcommand{\BBzero}{BB$_0$}
\newcommand{\BBthree}{BB$_3$}
\newcommand{\BBsix}{BB$_6$}

\begin{document}

\preprint{APS/123-QED}

\title{Generation of neutron Bessel beams}

\author{Owen Lailey} 
\email{oalailey@uwaterloo.ca}
\affiliation{Institute for Quantum Computing, University of Waterloo,  Waterloo, ON, Canada, N2L3G1}
\affiliation{Department of Physics and Astronomy, University of Waterloo, Waterloo, ON, Canada, N2L3G1}

\author{Dusan Sarenac}
\affiliation{Department of Physics, University at Buffalo, State University of New York, Buffalo, New York 14260, USA}

\author{Nir Emelianenko} 
\affiliation{Institute for Quantum Computing, University of Waterloo,  Waterloo, ON, Canada, N2L3G1}
\affiliation{Department of Physics and Astronomy, University of Waterloo, Waterloo, ON, Canada, N2L3G1}

\author{Charles W. Clark}
\affiliation{Joint Quantum Institute, National Institute of Standards and Technology and University of Maryland, College Park, Maryland 20742, USA}

\author{David G. Cory}
\affiliation{Institute for Quantum Computing, University of Waterloo,  Waterloo, ON, Canada, N2L3G1}
\affiliation{Department of Chemistry, University of Waterloo, Waterloo, ON, Canada, N2L3G1}

\author{Huseyin Ekinci} 
\affiliation{Institute for Quantum Computing, University of Waterloo,  Waterloo, ON, Canada, N2L3G1}
\affiliation{Department of Physics and Astronomy, University of Waterloo, Waterloo, ON, Canada, N2L3G1}

\author{Michael G. Huber}
\affiliation{National Institute of Standards and Technology, Gaithersburg, Maryland 20899, USA}

\author{Jeremy W. Paster}
\affiliation{National Institute of Standards and Technology, Gaithersburg, Maryland 20899, USA}

 \author{Anna Sokolova}
\affiliation{Australian Centre for Neutron Scattering, ANSTO, Lucas Heights, New South Wales, 2234, Australia}

 \author{Diego Alba Venero}
\affiliation{ISIS Neutron and Muon Source, Rutherford Appleton Laboratory, Harwell Oxford, Didcot, UK}

\author{Dmitry A. Pushin}
\email{dmitry.pushin@uwaterloo.ca}
\affiliation{Institute for Quantum Computing, University of Waterloo,  Waterloo, ON, Canada, N2L3G1}
\affiliation{Department of Physics and Astronomy, University of Waterloo, Waterloo, ON, Canada, N2L3G1}

\date{\today}

% \pacs{Valid PACS appear here}

\begin{abstract}
Matter waves can exhibit structured wavefields analogous to those of light, offering new means of controlling the spatial and angular-momentum degrees of freedom of massive particles. Bessel beams are a fundamental class of structured waves that exhibit non-diffracting propagation, self-healing, and well-defined orbital angular momentum (OAM). Here, we experimentally realize neutron Bessel beams by imprinting a nanoscale conical phase gradient onto a neutron wavefront. By incorporating azimuthal phase gradients, we generate zeroth-, third-, and sixth-order Bessel beams, demonstrating direct control of both the OAM and radial structure of a massive particle. We also propose a neutron Bessel-grating interferometer, complementary to established optical and electron interferometric methods, in which varying the radial grating period by approximately 100~nm yields a controlled Gouy phase shift of $\pi$. These results establish neutron Bessel beams as a new platform for diffractive neutron focusing and provide an experimental foundation for theoretical studies of Bessel beams in neutron scattering, decay, and OAM characterization.
\end{abstract}
\maketitle

\section{Introduction}
In 1987, Durnin \textit{et al.} experimentally demonstrated a remarkable class of propagation-invariant solutions to the Helmholtz equation: Bessel beams (BBs)~\cite{durnin1987exact, durnin1987diffraction}. Unlike conventional Gaussian beams, an ideal BB maintains its transverse intensity profile during propagation and can reconstruct after partial obstruction~\cite{bouchal1998self}. Furthermore, zeroth-order BBs possess the smallest central-spot diameter compared to conventional beams~\cite{mcgloin2005bessel}, while higher-order BBs carry well-defined orbital angular momentum (OAM)~\cite{volke2002orbital}. These properties have enabled applications including extended-depth-of-field imaging~\cite{planchon2011rapid, fahrbach2012propagation}, OAM transfer~\cite{garces2003observation}, optical micromanipulation~\cite{arlt2001optical, garces2002simultaneous}, and high-aspect-ratio laser materials processing~\cite{duocastella2012bessel}.

While BBs have been extensively studied in optical systems~\cite{rubinsztein2016roadmap}, their subsequent realization with matter waves~\cite{bliokh2023roadmap}, including electrons~\cite{grillo2014generation} and atomic Bose--Einstein condensates~\cite{ryu2014creation}, demonstrated that these propagation-invariant solutions are a universal feature of wave mechanics. Several theoretical studies have directly considered free-space propagating neutron Bessel beams, and their relevance in multiple neutron scattering experiments~\cite{afanasev2019schwinger, afanasev2021elastic, sherwin2022scattering}, neutron decay~\cite{pavlov2025angular}, and neutron OAM characterization~\cite{maksimov2025diffraction}. Realizing such beams with neutrons, however, presents unique experimental challenges arising from limited transverse coherence, low available flux, and finite detector spatial resolution. These constraints necessitate different approaches to wavefront engineering than those used for optical or charged-particle beams.

Here, we apply the nanoscale phase-grating array approach developed for generating helical neutron waves~\cite{sarenac2022experimental} to imprint a conical phase gradient onto the neutron wavefront, generating free-space propagating neutron BBs. We demonstrate control over both the radial structure and OAM of the neutron wave by generating zeroth-, third-, and sixth-order BBs. For the zeroth-order BB propagated into the far-field, we measure a central intensity spot-size FWHM that is a factor of $2\times$ smaller than that of a Gaussian beam under equivalent experimental conditions. Together with vortex~\cite{sarenac2022experimental} and Airy~\cite{sarenac2025generation} beams, this work completes the three fundamental classes of structured neutron states. The conical structure of neutron Bessel beams also provides a phase-grating platform for measuring the Gouy phase, a fundamental phase anomaly described by the uncertainty principle~\cite{feng2001physical}, which has been observed for optical Bessel beams~\cite{martelli2010gouy} and structured electron beams~\cite{guzzinati2013observation}. Finally, neutron Bessel beams offer a route to shaping and focusing neutron waves without refractive optics, which are limited by near-unity refractive indices~\cite{Sam, Sears}.
\raggedbottom

\begin{figure*}
    \centering
    \includegraphics[width=1\linewidth]{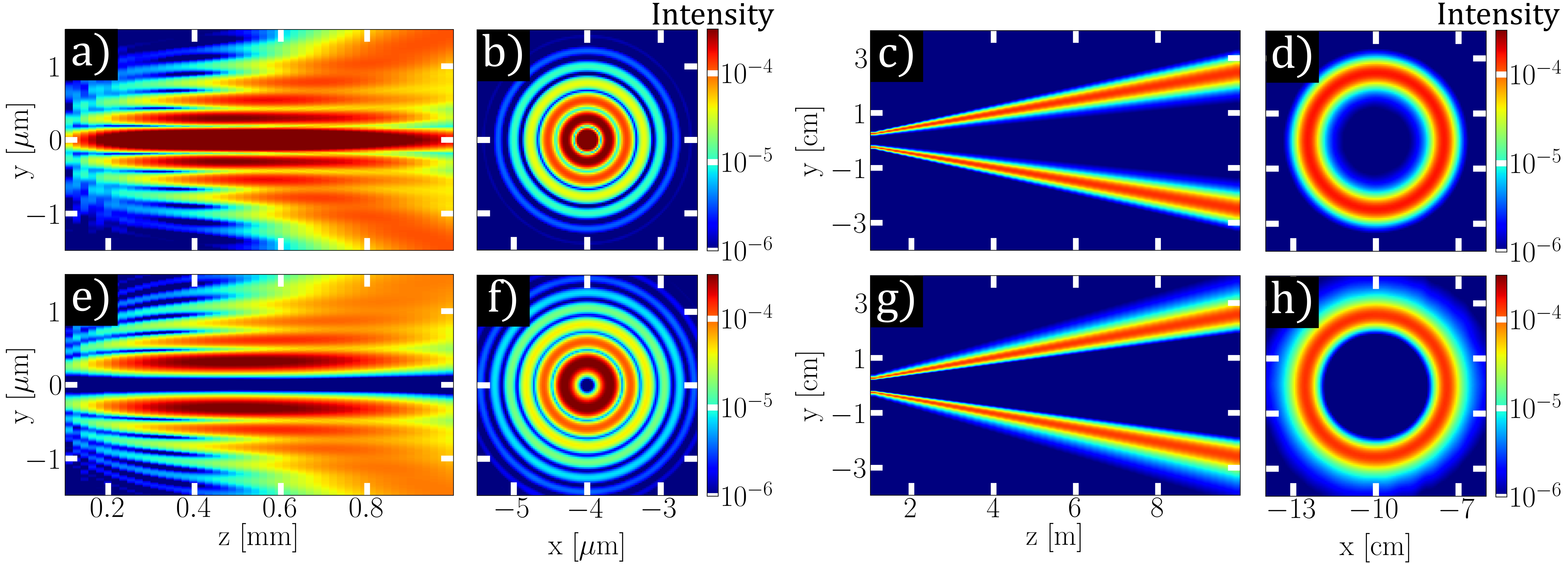}
    \caption{Simulated near-field and far-field propagation of Bessel beams generated by two phase gratings: a $q=0$ grating producing a zeroth-order Bessel beam (\BBzero{}, top row) and a $q=3$ grating producing a third-order Bessel beam (\BBthree{}, bottom row). Simulation parameters are $p=120$~nm, $\rho=480$~nm, $R=10~\mu$m, $\lambda=12$~\AA{}, and $\sigma=1~\mu$m. (a) Near-field intensity carpet of \BBzero{} in the first diffraction order. Consistent with Eq.~\ref{eq:nondiff}, the beam remains non-diffracting until $z\approx0.4$~mm, beyond which it begins to expand. (b) Transverse ($xy$) intensity profile at $z=0.4$~mm, showing a bright central core surrounded by concentric rings. (c) Far-field intensity carpet up to $z=10$~m and (d) the corresponding transverse profile at $z=10$~m, where the beam forms a ring. (e--h) Same as (a--d) for \BBthree{}, illustrating the effect of OAM. In the near field (e,f), the bright central core of \BBzero{} is replaced by a zero-intensity core arising from the on-axis phase singularity. In the far field (g,h), the ring has a larger radius than for \BBzero{} from carrying nonzero OAM.}
    \label{fig1}
\end{figure*}

\section{Bessel Beam Generation}
Several review articles have presented the theory, experimental generation, and applications of Bessel beams~\cite{khonina2020bessel, rao2024conceptual}. Here, we outline the basic principles underlying the generation and propagation of neutron Bessel beams implemented in this work.

\textit{Theory:} A Bessel beam is a cylindrical-coordinate solution of the time-dependent Schrödinger equation for a free neutron, which can be written as
\begin{equation}
\psi(r,\phi,z,t)
=
J_{\ell}(k_r r)
e^{i\ell\phi}
e^{ik_z z}
e^{-i\omega t},
\label{eq:schro}
\end{equation}
where $J_{\ell}(k_r r)$ is the Bessel function of the first kind of order $\ell$, $k_r$ and $k_z$ are the transverse and longitudinal components of the neutron wavevector, respectively, and $\omega = \hbar k^2/2m_n$, where $k^2 = k_r^2 + k_z^2$ and $m_n$ is the neutron mass. The integer $\ell$ denotes the OAM mode of the beam, with each neutron carrying OAM of $\ell\hbar$ about the propagation axis.

The transverse profile of a Bessel beam can be understood as the coherent superposition of plane waves whose wavevectors lie on a cone, with equal transverse wavevector magnitude ($k_r$) and uniformly distributed azimuthal angles. This superposition gives rise to the Bessel function $J_{\ell}(k_r r)$ in Eq.~\ref{eq:schro}, which can be expressed as~\cite{NIST:DLMF}
\begin{equation}
\begin{split}
J_\ell(k_r r) = \frac{i^{-\ell}}{2\pi}\int_0^{2\pi}
& \exp[{i\ell(\phi'-\phi)]} \\
&\times \exp\!\left[ik_r\left(x\cos\phi' + y\sin\phi'\right)\right] d\phi'.
\end{split}
\label{eq:bessel_superposition}
\end{equation}

\textit{Propagation dynamics:}
A Bessel beam can be generated by imprinting a conical phase profile onto an incident wavefront using a phase grating. A general Bessel phase profile is given by
\begin{equation}
\Phi(x,y) = \frac{\Delta\beta}{2}\bigg[\text{sgn}\left(\sin\left[\frac{2\pi}{p}x+\frac{2\pi}{\rho}r+q\phi\right]\right)+1\bigg],
\label{eq:grating}
\end{equation}
where $p$ is the period of the linear grating, $\rho$ is the radial period, and $q$ is the topological charge of the grating. The phase shift acquired by the neutron wavefunction relative to free space is set by the grating height $h$ according to $\Delta\beta = N b_c \lambda h$, where $Nb_c$ is the neutron scattering length density of the grating material and $\lambda$ is the neutron wavelength. The Bessel-beam order $\ell$, and hence the OAM mode, is determined by the grating topological charge and the diffraction order $m$ according to $\ell=mq$, such that neutrons in the $m$th diffraction order carry an OAM of $mq\hbar$. The sign of $mq$ determines the handedness of the OAM winding about the propagation axis.

The linear grating component spatially separates the diffraction orders at the detector, with the first diffraction orders propagating at an angle $\alpha\approx\lambda/p$ relative to the zeroth order. The conical phase fringes set the cone angle $\theta\approx\lambda/\rho$ of the Bessel beams. In the far field at a propagation distance $z$, the $m=+1$ diffraction order for $q=0$ is centered at horizontal position $x\approx z\alpha$, with a Bessel ring radius of $r\approx z\theta$. Thus, the linear and radial grating periods determine the diffraction angle and the transverse wavevector, respectively, while the azimuthal phase winding determines the quantized OAM $\ell\hbar$. The resulting radial intensity profile depends on both the transverse wavevector and the OAM order $\ell$.

Figure~\ref{fig1} illustrates the near-field and far-field propagation of neutron Bessel beams generated by gratings with $q=0$ (top row) and $q=3$ (bottom row), producing zeroth- and third-order Bessel beams that differ by whether the beam carries OAM. The simulations consider a Gaussian neutron wavepacket with transverse extent $\sigma$ incident on a phase grating described by Eq.~\ref{eq:grating}, with $p=120$~nm, $\rho=480$~nm, $R=10~\mu$m, $\lambda=12$~\AA{}, and $\sigma=1~\mu$m. Details of the simulation method are provided in Refs.~\cite{sarenac2024small, sarenac2026probing}.

When the incident wavepacket uniformly illuminates the grating element ($\sigma \gg R$), the maximum propagation distance over which the Bessel beam remains non-diffracting is
\begin{equation}
z_{\text{max}} = \frac{R}{\tan\theta},
\label{eq:nondiff}
\end{equation}
where $R$ is the radius of the diffractive element and $\theta = \lambda/\rho$ is the cone angle. In the opposite limit ($R \gg \sigma$), the transverse extent of the wavepacket sets the effective aperture, and $R$ in Eq.~\ref{eq:nondiff} is replaced by $\sigma$. The simulations in Fig.~\ref{fig1} fall in this second regime ($R = 10\sigma$), giving $z_{\text{max}} \approx \sigma/\tan\theta \approx 0.4$~mm, and both beams remain non-diffracting up to this distance before beginning to spread (Figs.~\ref{fig1}a,e).

Within the non-diffracting region, both beams exhibit the multi-ring transverse profile characteristic of Bessel beams (Figs.~\ref{fig1}b,f), but the zeroth-order beam has a bright, narrow central core, whereas the beam carrying OAM has a zero-intensity core due to its on-axis phase singularity. Beyond $z_{\text{max}}$, the beams expand linearly (Figs.~\ref{fig1}c,g) and both form ring-shaped profiles in the far field (Figs.~\ref{fig1}d,h), with the ring radius increasing in the nonzero OAM case. The interplay between $R$, $\sigma$, and $\ell$ thus governs both the propagation dynamics and the transverse intensity profiles of the Bessel beams~\cite{overfelt1991comparison}.\\

\begin{figure}
    \centering
    \includegraphics[width=0.85\linewidth]{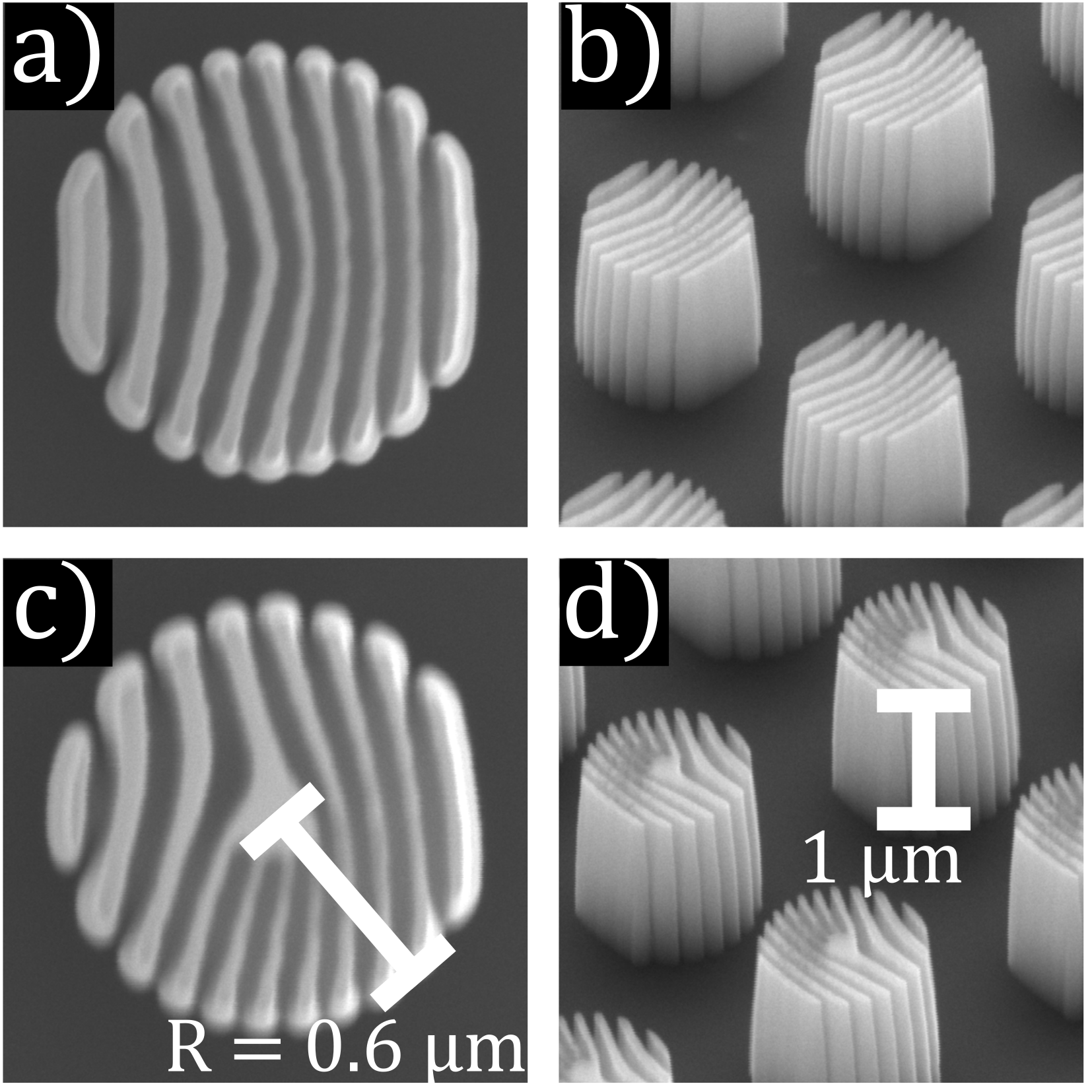}
    \caption{SEM images of the Bessel beam phase gratings used in this work. (a) A single grating element of the zeroth-order Bessel grating with $q=0$ (Eq.~\ref{eq:grating}). (b) The \BBzero{} phase-grating array viewed at a $45^\circ$ angle. (c,d) Corresponding images of the $q=3$ grating, which produces the \BBthree{} and \BBsix{} neutron Bessel beams. The grating element has a radius of $R=0.6~\mu$m.}
    \label{fig2}
\end{figure}

\textit{Experimental Implementation:} We implemented this method of Bessel beam generation by fabricating nanoscale conical phase structures in silicon wafers, with a linear period of $p=120~$nm and a radial period of $\rho=480~$nm. Scanning electron microscopy (SEM) images of individual grating elements are shown in Fig.~\ref{fig2}a for $q=0$ and Fig.~\ref{fig2}c for $q=3$, and the corresponding array structures, viewed at a $45^\circ$ angle, are shown in Fig.~\ref{fig2}b,d. Each grating element has a radius of approximately $R=0.6~\mu$m, chosen to be comparable to the transverse neutron coherence length~\cite{sarenac2026probing}, and the elements are repeated millions of times in an array to increase the overall diffraction efficiency. Since neutrons in the $m$th diffraction order carry an OAM of $mq\hbar$, the $q=3$ grating produces third-order Bessel beams (\BBthree{}) in the $m=\pm1$ diffraction orders and sixth-order Bessel beams (\BBsix{}) in the $m=\pm2$ diffraction orders. For the fabricated gratings, the non-diffraction region is confined to a short distance directly after the grating, $z_{\text{max}}\approx0.24$~mm for $\lambda = 12~$\AA.

\begin{table}[!b]
\centering
\renewcommand{\arraystretch}{1.3}
\setlength{\tabcolsep}{6pt}

\begin{tabular}{||c||c|c|c||}
\hline\hline
\textbf{Facility}      & ISIS   & ISIS   & ANSTO \\
\hline
\textbf{Instrument}    & ZOOM   & ZOOM   & BILBY \\
\hline\hline
\textbf{Measurement}   & \BBzero{} \& \BBthree{} & \BBsix{} & \BBthree{} \\
\hline
\textbf{$z$ [m]}       & 9.6        & 9.6       & 18.4 \\
\hline
\textbf{$\lambda$ [\AA]} & 10.5--14.5 & 6.5--14.5 & 12--19 \\
\hline
\textbf{$A_1$ [mm]}    & 6          & 6         & 20 \\
\hline
\textbf{$L_1$ [m]}     & 8          & 8         & 17.1 \\
\hline
\textbf{$A_2$ [mm]}    & 6          & 6         & 5 \\
\hline\hline
\end{tabular}
\caption{Experimental configurations for the measurements in this work. Measurements of the \BBzero{}, \BBthree{}, and \BBsix{} Bessel beams were performed at ISIS/ZOOM and ANSTO/Bilby. Here $z$ is the grating-to-detector distance, $\lambda$ is the wavelength range, and $A_1$, $L_1$, and $A_2$ are the source slit diamater, collimation distance, and sample slit diameter, respectively.}
\label{tab}
\end{table}

 \begin{figure*}
    \centering
    \includegraphics[width=1\linewidth]{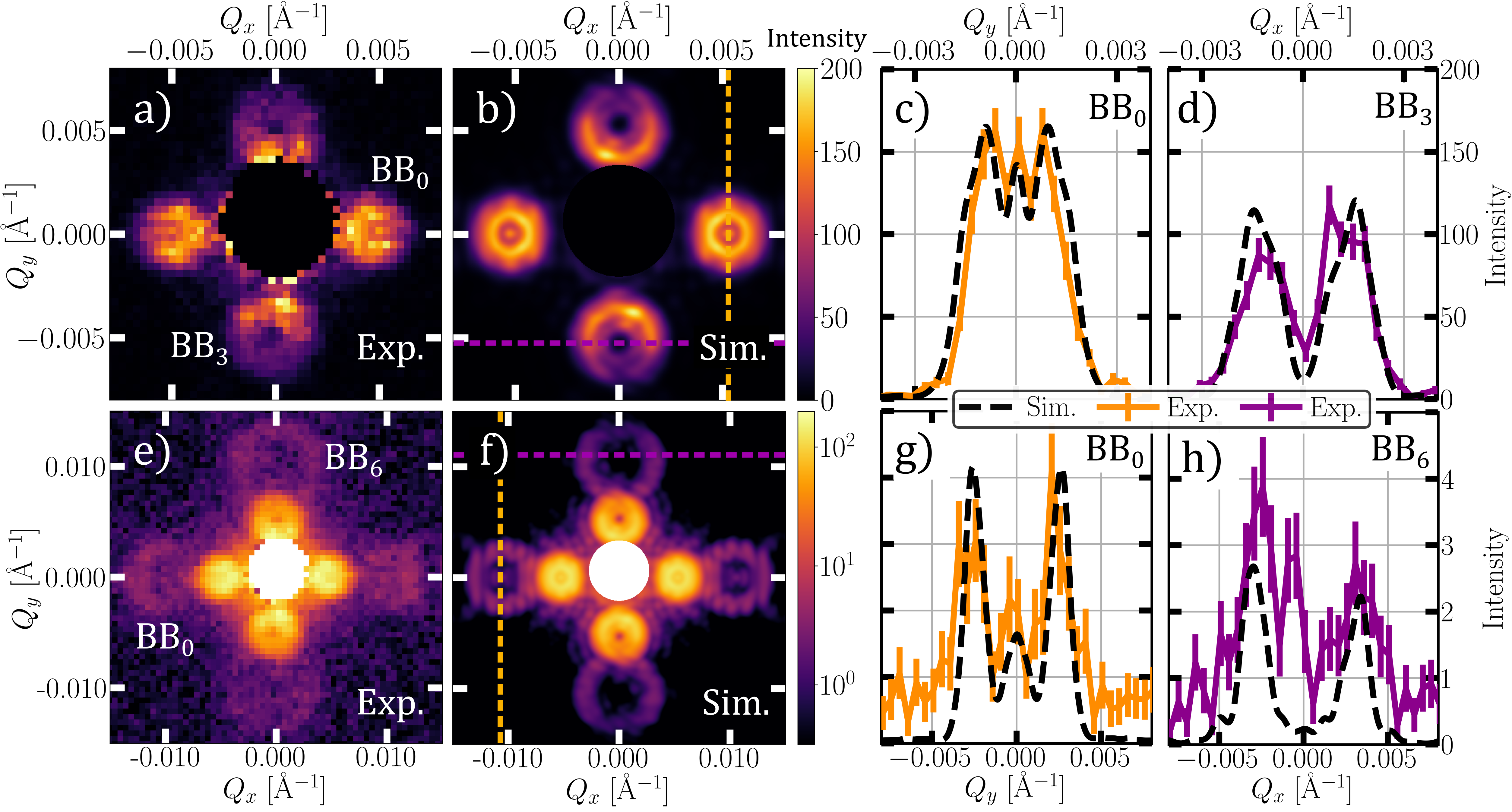}
    \caption{(a,b) Experimental and simulated far-field SANS intensity profiles of the zeroth order Bessel beam (\BBzero, horizontal diffraction) and the third order Bessel beam (\BBthree, vertical diffraction) at the 2D position sensitive neutron detector. As observed in both experiment and simulation, the \BBzero~possesses a bright central spot surrounded by a broader ring, whereas \BBthree~displays the zero intensity center from carrying OAM $\ell = 3$. (c,d) Experimental and simulated intensity slices along the \BBzero~and \BBthree~first diffraction orders, indicated by the orange and purple lines in (b). (e,f) Experimental and simulated far-field SANS intensity profiles in log-scale to demonstrate higher diffraction orders, such as the sixth order Bessel beam (\BBsix, vertical diffraction). (g,h) Transverse slices are in good agreement with simulation. The experiment exhibits some asymmetry between positive and negative diffraction orders, which may arise from slight tilts of the wafers about the vertical and horizontal axes, and which may explain why the simulated \BBsix~is slightly less intense. In both experiment and simulation the direct beam is masked out to emphasize the diffracted orders. The experimental configurations for the top row and bottom row of images correspond to columns 2 and 3 of Table~\ref{tab}, respectively.}
    \label{fig:intensity}
\end{figure*}

Using these gratings, we performed neutron scattering experiments on the ZOOM small-angle neutron scattering (SANS) instrument at the ISIS Neutron and Muon Source and on the Bilby SANS instrument~\cite{sokolova2019performance} at the Australian Centre for Neutron Scattering (ACNS). An overview of the measurements and the relevant experimental parameters for both instruments is given in Table~\ref{tab}. The standard SANS setup consists of a collimation region defined by the source slit with diameter $A_1$, the collimation distance $L_1$, and the sample slit with diameter $A_2$, followed by the sample position, occupied here by the silicon phase gratings. The neutrons then propagate a distance $z$ to a position-sensitive $^3$He neutron detector with typical pixel sizes of $\sim$3--8~mm.

Two-dimensional data reduction was performed for all measurements using the Mantid framework~\cite{arnold2014mantid}. The ZOOM configuration was optimized for measurement of multiple diffraction orders simultaneously, while Bilby was used in a high-resolution configuration, taking advantage of its long rear-detector distance and adjustable wavelength resolution, which we set to $\Delta\lambda/\lambda = 5.6~\%$. Only Bilby's rear detector was used. Bilby data were reduced using dedicated scripts~\cite{MantidBilby}, with the wavelength range restricted to 12--19~\AA{} to maximize signal while excluding shorter wavelengths, where the angular resolution and detector pixelation are less favorable. 

\section{Results}
Shown in Fig.~\ref{fig:intensity} are the experimentally generated \BBzero, \BBthree, and \BBsix~far-field intensity profiles, alongside simulations. The measurement configurations for rows 1 and 2 in Fig.~\ref{fig:intensity} correspond to columns 2 and 3 of Table~\ref{tab}, respectively. To simultaneously measure both $q=0$ and $q=3$ BB gratings, the BB gratings were stacked orthogonally, utilizing a method previously proposed and implemented with fork-dislocation phase-gratings~\cite{lailey2026multimode, sarenac2026experimental}. As seen in Fig.~\ref{fig:intensity}a, the $q=0$ grating is arranged for horizontal diffraction, and the $q=3$ grating is perpendicular, producing the vertical diffraction orders. The black or white central masks obstruct the direct beam in both simulations and experiments. 

We directly compare transverse intensity slices between simulation and experiment. The orange and purple lines represent the experiment slices while the black dotted line is the simulation. All simulated intensity profiles are in good agreement with experiment for simulated grating heights of $h = 1.2~\mu$m for the $q=0$ grating and $h = 1.3~\mu$m for the $q=3$ grating. The diffraction efficiency of the gratings is proportional to $h^2$ for small phase shifts ($\Delta\beta$ in Eq.~\ref{eq:grating}).

\begin{table}[!b]
\centering
\caption{Neutron Bessel beam quality, quantified by the fringe contrast $C$ (Eq.~\ref{eq:contrast}) for each Bessel-beam order. The uncertainties are statistical and represent one standard deviation.}
\label{tab:BB_contrast}
\renewcommand{\arraystretch}{1.3}
\setlength{\tabcolsep}{12pt}

\begin{tabular}{||c||c||}
\hline\hline
\textbf{BB order} & \textbf{Contrast} \\
\hline\hline
\BBzero{} ($m=1$) & $(15 \pm 7)~\%$ \\
\hline
\BBzero{} ($m=2$) & $(61 \pm 7)~\%$ \\
\hline
\BBthree{} & $(61 \pm 13)~\%$ \\
\hline
\BBsix{} & $(73 \pm 12)~\%$ \\
\hline\hline
\end{tabular}
\end{table}

The structure in the BB intensity profiles is a direct result of interference taking place between the waves propagating in a conical shape. We can therefore quantify the quality of the neutron BB production through visibility of the interference fringes~\cite{rao2024conceptual}. Comparing the bright ring intensities ($I_\text{max}$) to the dark fringes ($I_\text{min}$) for all profiles in Fig.~\ref{fig:intensity}, we compute the contrast
\begin{equation}
    C = \frac{I_\text{max} - I_\text{min}}{I_\text{max} + I_\text{min}}
    \label{eq:contrast}
\end{equation}
and summarize the results in Table~\ref{tab:BB_contrast}. The contrast of the \BBzero~in the first diffraction order was limited to $15\%$ due to detector spatial resolution constraints, while the \BBthree~and \BBsix~with the dark vortex centers have a production quality of over $60~\%$.

Notably, both simulated and experimental \BBzero~intensity profiles in Fig.~\ref{fig:intensity} illustrate a small central lobe. As illustrated in Fig.~\ref{fig1}a-h, this is expected behaviour for the near-field \BBzero~intensity profiles but in the far-field there is typically no central spot for any BB. This far-field central lobe appears since we are in the limit $R < \sigma$. In this aperture-edge limit, the resultant far-field intensity profile is effectively a convolution of the ideal ring pattern shown in Fig.~\ref{fig1}d, and the diffraction from the phase grating aperture edge in Fig.~\ref{fig2}a. This yields a central lobe in the far-field \BBzero~intensity profiles, which we see in both experiment and simulation in Fig.~\ref{fig:intensity}a,b.

Computing the FWHM for the first diffraction order \BBzero~we obtain a central spot FWHM of $7.8~\text{mm}~\pm~0.1~$mm, at $\lambda=12.5~$\AA. For comparison, we performed measurements with solely linear gratings at the exact same propagation distance and collimation settings. The linear gratings have been previously demonstrated in Refs.~\cite{sarenac2022experimental, sarenac2025generation}. The FWHM of the Gaussian diffraction spots in the first diffraction order is $15.5~\text{mm}~\pm~0.1~$mm. Under these experimental conditions, the neutron BBs have a central spot size that is $\sim2$ times smaller than conventional Gaussian beams.

 \begin{figure}
    \centering
    \includegraphics[width=1\linewidth]{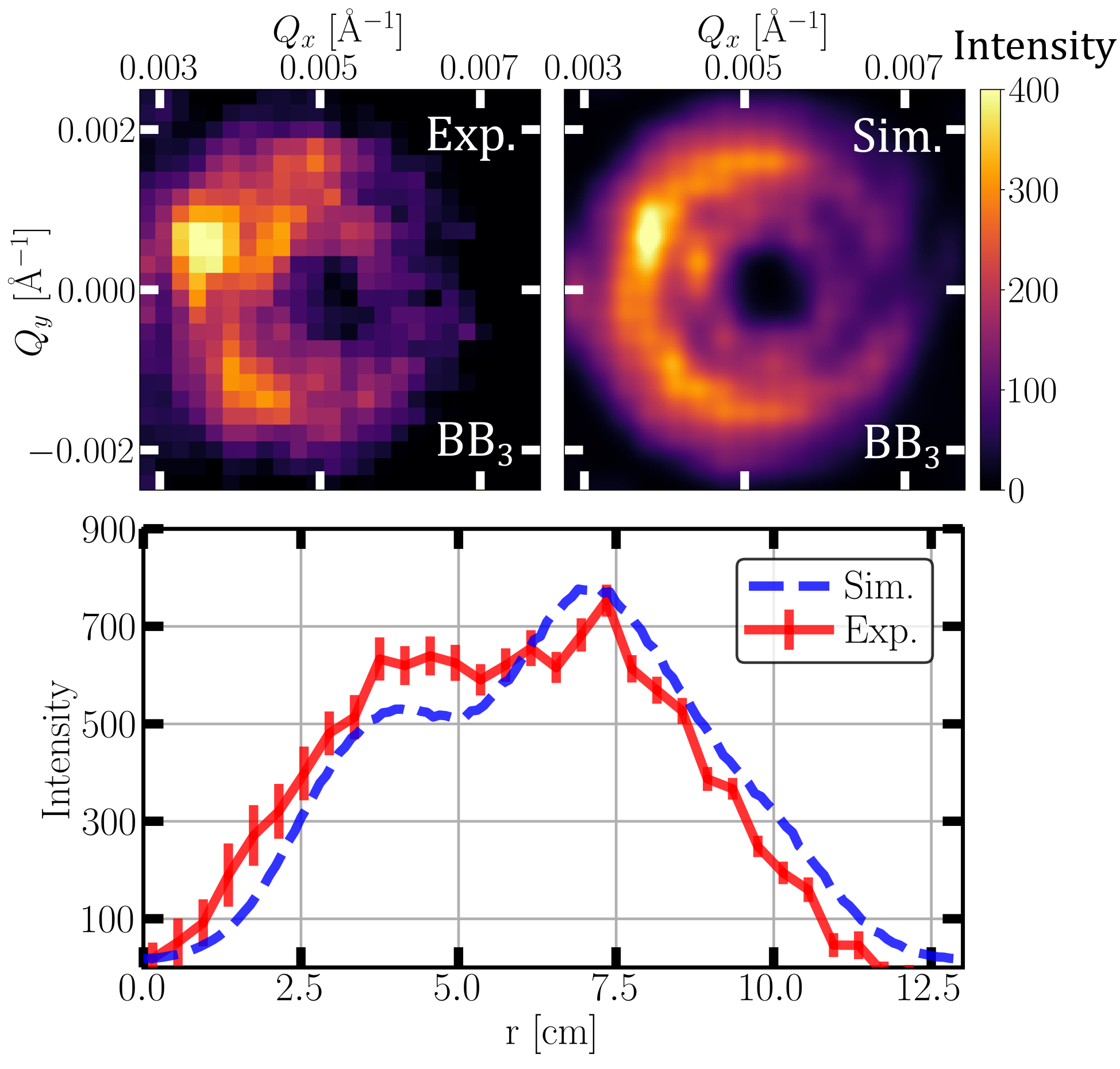}
    \caption{Experimental and simulated third-order Bessel beam \BBthree~measured using the higher angular resolution configuration in column 4 of Table~\ref{tab}, which resolves the more complex ring structure comprising the intensity profile. A dimmer inner ring is observed in both experiment and simulation, surrounded by a larger-radius, more intense outer ring. The bright spot on the outer ring is a direct consequence of the physical phase-grating structure. The corresponding azimuthally integrated radial intensity profiles show peaks at $r\sim4$~cm and $r\sim7$~cm for the inner and outer rings, respectively. The overall radial profile is substantially broader than that produced by a simple $\ell=3$ fork-dislocation phase-grating.}
    \label{fig:ansto}
\end{figure}

The \BBthree~profiles in Fig.~\ref{fig:intensity} exhibit a faint multi-ring structure, characterized by a bright lobe on the outer ring and a weaker inner ring. A similar structure is reproduced in our simulations. In the ideal limit of large $R$, a Bessel beam is expected to produce a single, narrow ring in the far field (Fig.~\ref{fig1}d,h). Once again, the additional structure observed here arises from the interplay between the finite values of $R$ and $\rho$ and the neutron's transverse extent.

To resolve these features more clearly, we generated \BBthree~using a higher angular resolution configuration, listed in column 4 of Table~\ref{tab}. Figure~\ref{fig:ansto} shows the $m=+1$ diffraction order from the $q=3$ grating, which produces \BBthree. The bright outer lobe is clearly resolved, while the weaker inner ring is also visible. The corresponding azimuthally integrated radial intensity profiles from experiment and simulation are shown below the \BBthree~images in Fig.~\ref{fig:ansto}. Both profiles exhibit a central intensity minimum at $r=0$, consistent with the phase singularity associated with the $\ell=3$ OAM mode, followed by a weak peak at approximately $r=4.0$~cm corresponding to the inner ring. At larger radius, a second, more intense peak corresponds to the bright outer ring, before the intensity decreases to a minimum beyond $r=10$~cm. In comparable experiments featuring small phase grating apertures, this wide ring structure distinguishes the \BBthree~profile from a conventional $\ell=3$ OAM mode~\cite{sarenac2022experimental}.

\section{Discussion and Applications}
The experimental realization of neutron Bessel beams establishes them as an accessible class of structured neutron wavefields and completes the demonstration of the three fundamental classes of structured neutron states: vortex (OAM)~\cite{sarenac2022experimental}, Airy~\cite{sarenac2025generation}, and Bessel beams, expanding the range of structured wavefields available for neutron optics. This development comes amid recent progress in neutron OAM research, including new approaches to OAM generation and characterization~\cite{geerits2025measuring, mckay2026topological, chahal2026generalized, geerits2026white, samiei2026generalized} and investigations of OAM-dependent nuclear interactions~\cite{sarenac2026experimental}. Beyond just vortex states, various theoretical studies have directly considered neutron Bessel beams for investigating scattering~\cite{afanasev2019schwinger, afanasev2021elastic, sherwin2022scattering}, OAM characterization~\cite{maksimov2025diffraction}, and neutron decay~\cite{pavlov2025angular}. More recently, theoretical work has investigated spin textures and Skyrmion-like structures in twisted fermion beams, including Bessel states~\cite{afanasev2026non}. 

Notably, the conical structure of neutron Bessel beams also provides access to the Gouy phase. This phase can be understood as a consequence of the uncertainty principle: confining a wave transversely spreads its transverse momentum and thereby reduces its longitudinal wavevector~\cite{feng2001physical}. In a Bessel beam, this confinement is built into the cone of plane waves itself. Because every plane wave component shares the same transverse wavevector $k_r$, the beam accumulates a Gouy phase that grows linearly with propagation distance and is independent of the OAM order $\ell$~\cite{martelli2010gouy, rao2024conceptual},
\begin{equation}
\phi_G = (k-k_z)\,z \approx \frac{k_r^2 z}{2k} = \frac{\pi \lambda z}{\rho^2}.
\label{eq:gouy}
\end{equation}
For neutrons, this confinement-induced phase has been observed by passing neutrons through narrow silicon channels inside a perfect-crystal interferometer~\cite{rauch2002measurement}. Bessel gratings offer a complementary free-space measurement, with the transverse momentum defined lithographically by $\rho$ rather than by material boundaries, and with a phase that accumulates at $\approx 16$~rad/mm for $\lambda = 12$~\AA{} and $\rho = 480$~nm.

\begin{figure}
    \centering
    \includegraphics[width=1\linewidth]{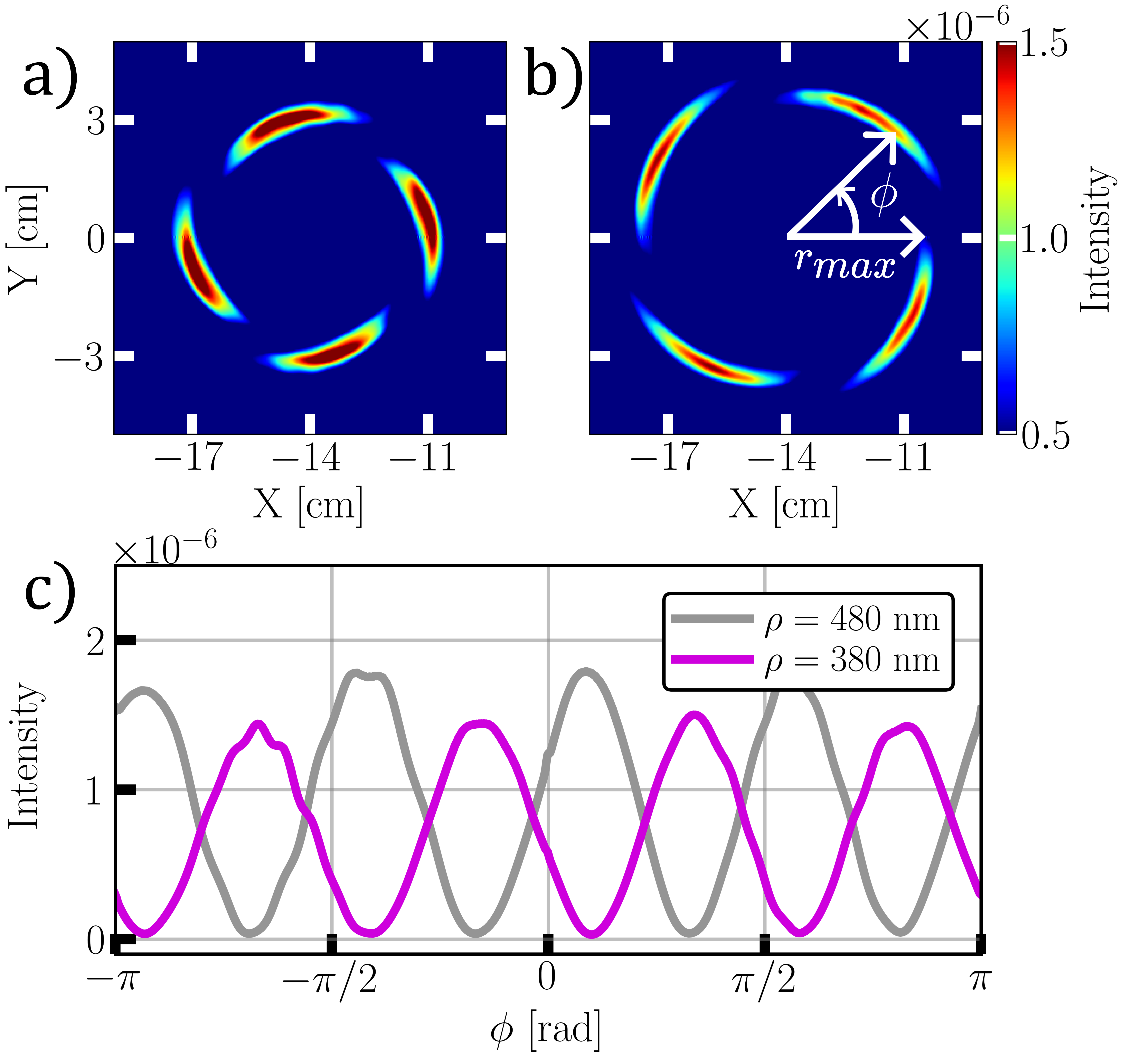}
    \caption{Simulated Gouy phase measurement with a neutron Bessel grating interferometer, consisting of $q=+2$ and $q=-2$ Bessel gratings on opposite faces of a silicon wafer of thickness $d=320~\mu$m, for $\lambda=12$~\AA. (a) Four-lobe far-field interference pattern for $\rho=480$~nm. (b) Reducing the radial period to $\rho=380$~nm increases the Gouy phase accumulated across the wafer by $\pi$ (Eq.~\ref{eq:gouy}), rigidly rotating the pattern by $\pi/\Delta\ell = \pi/4$~rad. The larger ring radius reflects the larger cone angle $\theta\approx\lambda/\rho$. (c) Azimuthal intensity profiles taken at the radius of maximum intensity slice ($r_{max})$ in (a) and (b), showing the $\pi$ phase shift. The simulations consider a single grating element on each face, with $R=2~\mu$m, $p=100$~nm, $z = 12$ m and $\sigma=10~\mu$m.}
    \label{fig:gouy}
\end{figure}

Interferometric measurements of the Bessel-beam Gouy phase are well established in optics~\cite{martelli2010gouy, rop2012measuring}, and Gouy-phase effects have been observed with structured electron beams~\cite{guzzinati2013observation, grillo2017observation}. These techniques translate naturally to neutron optics. For example, in a Bessel-grating variant of the neutron OAM interferometer~\cite{sarenac2024small}, arrays of $q=+2$ and $q=-2$ Bessel gratings with identical $\rho$ can be patterned on opposite faces of a silicon wafer of thickness $d$. The two primary interfering paths differ by the Gouy phase $\phi_G$ of Eq.~\ref{eq:gouy} with $z=d$, which rotates the four-fold petal interference pattern in the far field by $\phi_G/\Delta\ell$, where here $\Delta\ell=4$. Because $\phi_G \propto 1/\rho^2$, this rotation is set by the grating design: for $d = 320~\mu$m and $\lambda = 12$~\AA, changing $\rho$ from $480$~nm to $380$~nm shifts the Gouy phase by $\pi$, producing a clearly resolved petal rotation (Fig.~\ref{fig:gouy}). Such a measurement would provide a lithographically tunable, $\ell$-independent propagation phase for neutron interferometry, against which OAM-dependent phase shifts from materials and nuclear interactions could be isolated.

Beyond their OAM and radial features, neutron Bessel beams uniquely combine properties that are otherwise distributed across different classes of structured waves. The combination of a narrow central lobe and its propagation invariance over an extended non-diffracting region is particularly valuable for neutrons, for which focusing optics remain a central challenge: the neutron refractive index of common materials differs from unity by only $\sim 10^{-5}$--$10^{-6}$, so conventional refractive lenses are weak and must be stacked to achieve appreciable focusing~\cite{Sam, Sears}. A Bessel grating instead acts as a diffractive axicon, producing an extended focal line whose central lobe is set lithographically and is maintained over the entire non-diffracting region. By optimizing the grating parameters of this work, in particular the element radius $R$ and radial period $\rho$ in Eq.~\ref{eq:nondiff}, samples placed within the non-diffracting region could exploit the propagation-invariant and self-healing properties of neutron Bessel beams to investigate neutron-matter interactions with structured incident wavefronts. Finally, Bessel gratings are naturally complementary to emerging diffractive neutron optics. Neutron Fresnel lenses~\cite{dhanalakshmi2025diffractive, dhanalakshmi2026achromatic} could transform the generated Bessel beams into perfect vortices, whose ring radius is ideally independent of OAM~\cite{Ostrovsky:13}. Such beams would provide a platform for isolating OAM-dependent scattering effects, since the transverse extent of different OAM modes could ideally be made equal.

\section{Conclusion}
This work presents the first experimental realization of neutron Bessel beams and demonstrates control over their OAM and radial structure. By varying the grating topological charge, we generate zeroth-, third-, and sixth-order Bessel beams. The far-field intensity profile generated from the zeroth-order Bessel beam grating exhibits a central spot approximately $2\times$ smaller than the corresponding Gaussian diffraction order measured under identical conditions, demonstrating the enhanced transverse localization afforded by the Bessel-beam structure. Higher-order neutron Bessel beams are generated with a contrast quality of over $60~\%$. These results enable several applications, including measurement of the neutron Gouy phase, which can be shifted by $\pi$ through a $\sim100$~nm change in $\rho$, structured probes of OAM-dependent scattering and nuclear interactions, and diffractive neutron focusing without refractive optics.

\section*{Acknowledgments}
This work was supported by the Canadian Excellence Research Chairs (CERC) program, the Natural Sciences and Engineering Research Council of Canada (NSERC) Discovery program, the NSERC Canada Graduate Scholarships programs (CGS-M and PGS-D), Collaborative Research and Training Experience (CREATE) program, the Canada  First  Research  Excellence  Fund  (CFREF), and the US Department of Energy, Office of Nuclear Physics, under Interagency Agreement 89243019SSC000025. Experiments at the ISIS Neutron and Muon Source were supported by beamtime allocation RB2520246 from the Science and Technology Facilities Council. We acknowledge the support of the Australian Centre for Neutron Scattering (ACNS) and the Australian Government through the National Collaborative Research Infrastructure Strategy, in supporting the neutron research infrastructure used in this work via ACNS P21610.

\bibliographystyle{ieeetr}
\bibliography{mybib.bib}

\end{document}